\documentclass{entcs} \usepackage{prentcsmacro}
\usepackage{graphicx}

\newcommand{\be}{\begin{equation}}
\newcommand{\ee}{\end{equation}}
\newcommand{\bq}{\begin{eqnarray}}
\newcommand{\eq}{\end{eqnarray}}

\newcommand{\ket}[1]{\left | \, #1 \right\rangle}
\newcommand{\bra}[1]{\left \langle #1 \, \right |}

\def\lastname{Wootton and Pachos}
\begin{document}
\begin{frontmatter}
  \title{Universal Quantum Computation with Abelian Anyon Models}
 \author{James R. Wootton\thanksref{ALL}\thanksref{myemail}}
  \address{School of Physics and Astronomy\\ University of Leeds\\
    Leeds, UK}
 \author{Jiannis K. Pachos\thanksref{ALL} \thanksref{coemail}}
  \address{School of Physics and Astronomy\\ University of Leeds\\
    Leeds, UK}
 \thanks[ALL]{Thanks to Gavin K. Brennen, Ville Lahtinen and Zhenghan Wang for inspiring conversations.}
 \thanks[myemail]{Email:
    \href{mailto:phyjrw@leeds.ac.uk} {\texttt{\normalshape
        phyjrw@leeds.ac.uk}}}
 \thanks[coemail]{Email:
    \href{mailto:j.k.pachos@leeds.ac.uk} {\texttt{\normalshape
        j.k.pachos@leeds.ac.uk}}}
\begin{abstract} 
  We consider topological quantum memories for a general class of abelian anyon models defined on spin lattices. These are non-universal for quantum computation when restricting to topological operations alone, such as braiding and fusion. The effects of additional non-topological operations, such as spin measurements, are studied. These are shown to allow universal quantum computation, while still utilizing topological protection. Our work gives an insight into the relation between abelian models and their non-abelian counterparts.
\end{abstract}


\end{frontmatter}

\section{Introduction}\label{intro}

Anyons are quasiparticles that can exist on two-dimensional systems \cite{wilczek,preskill}. Their non-local behaviour is an interesting topic for foundational research in quantum physics \cite{levin,wen}. Importantly, one may employ anyons for fault-tolerant quantum computation \cite{kitaev,dennis,brennenpachos}. The so-called non-abelian anyon models receive most attention, since they possess straightforward means to store, protect and manipulate quantum information. Even so, proposals have been made to realize universal quantum computation using an abelian model, the toric code \cite{lloyd,pachos,raussendorf}. We extend the latter scheme to a wider class of abelian models, proving universality and fault-tolerance. This provides fresh insight into the power of abelian anyons in relation to their non-abelian counterparts. Physical realizations of abelian anyons are simpler than those for non-abelian anyons, giving an experimental motivation for our work.

We employ Kitaev's quantum double models \cite{kitaev}, expressed as stabilizer codes \cite{gottesman}. By analysing the outcome of stabilizer measurements, known as the syndrome, the type of error may be determined and corrected. The probability of the syndrome giving misleading results is suppressed by certain parameters of the code. We specifically consider models based upon the cyclic group of $d$ elements, $Z_d$, realized on a square lattice with a $d$-level spin on each edge. The elements $g \in Z_d$ are used to label basis states of $d$-dimensional spins, for which the generalised Pauli operators are defined,
\be
\sigma^x = \sum_{g \in Z_d} \ket{g+1}\bra{g}, \,\,\, \sigma^z = \sum_{g \in Z_d} \omega^g \ket{g}\bra{g}.
\ee
Here $\omega = e^{i 2 \pi / d}$. These satisfy the commutation relation $\sigma^z \sigma^x = \omega \sigma^x \sigma^z$. The eigenbasis and eigenvalues of the $\sigma^z$ operation are obvious. The eigenstates of $\sigma^x$ are those of the Fourier transform basis,
\be
\ket{\tilde g} = \frac{1}{\sqrt{d}} \sum_{h \in Z_d} \omega^{gh} \ket{h},
\ee
with corresponding eigenvalues $\omega^{-j}$. To rotate between these two bases, we use the Fourier transform,
\be
F = \sum_{g \in Z_d} \ket{\tilde g}\bra{g} = \frac{1}{\sqrt{d}} \sum_{g,h \in Z_d} \omega^{gh} \ket{h}\bra{g},
\ee
which has the properties $F^2\ket{g}=\ket{-g}$, $F^3 = F^{\dagger}$ and $F^4 = I$.

\begin{figure}[ht]
\begin{center}
{\includegraphics[width=10cm]{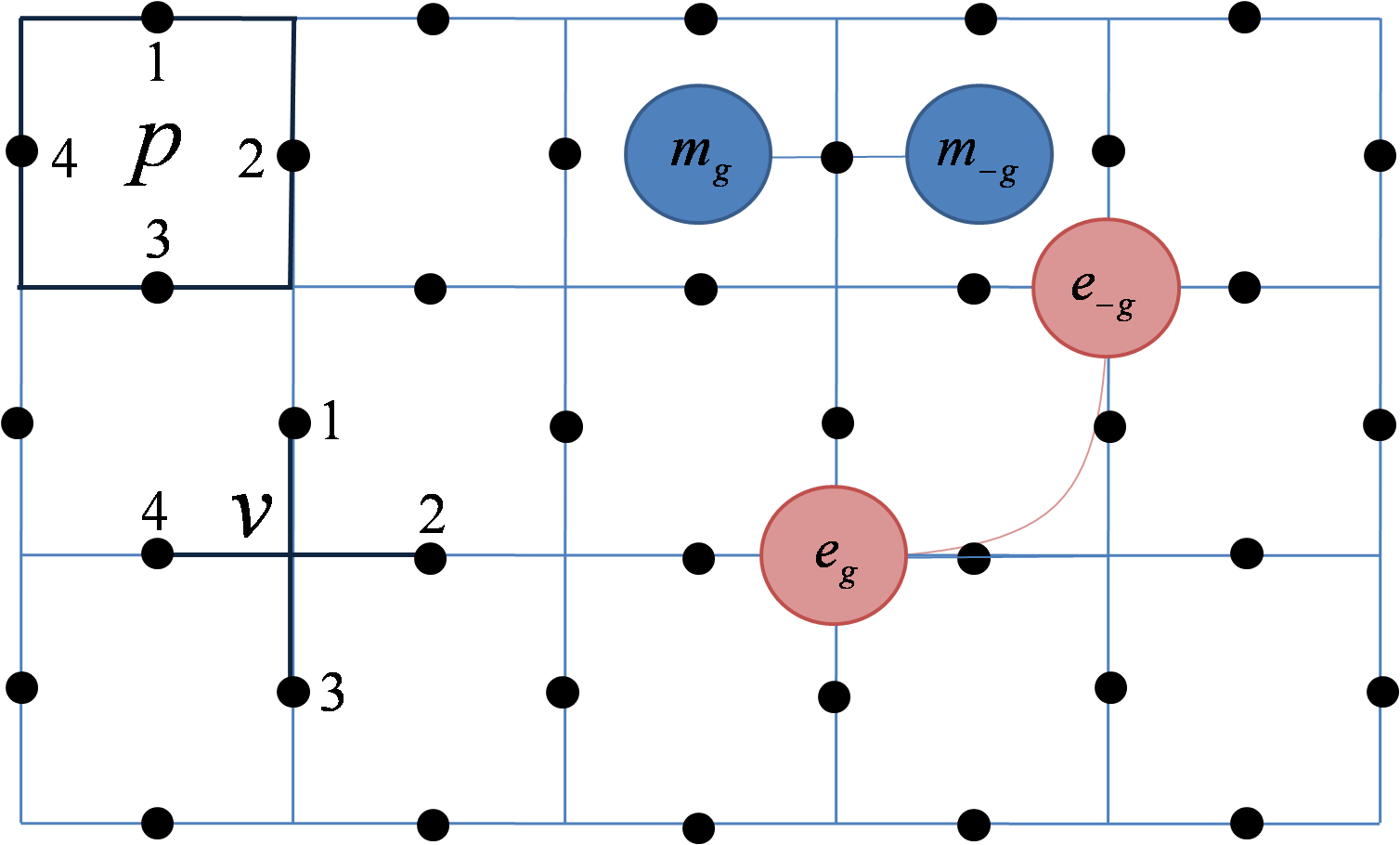} }
\caption{\label{fig1}The models are realized on a square lattice with a $d$-level spin on each edge. The numbering of spins around a plaquette, $p$, and vertex, $v$, is shown. Anyons $e_g$ reside on vertices. These may be created in pairs by single spin operations, and moved away from each other by chains of these operations along a string of spins. Anyons $m_g$ reside on plaquettes, and may be created and moved similarly.}
\end{center}
\end{figure}

The stabilizers of the quantum double models are defined on the four spins around each vertex, $v$, and plaquette, $p$,
\be
A(v) = \sigma^x_1{}^{\dagger} \sigma^x_2{}^{\dagger} \sigma^x_3 \sigma^x_4, \,\,\,
B(p) = \sigma^z_1{}^{\dagger} \sigma^z_2 \sigma^z_3 \sigma^z_4{}^{\dagger},
\ee
where the numbering proceeds clockwise from the top-most spin (Fig. \ref{fig1}). These have eigenvalues $\omega^g=e^{i 2 \pi g/d}$ for each $g \in Z_d$. No anyon is associated with a vertex or plaquette if $A(v) \ket\psi=B(p) \ket\psi=\ket\psi$. An anyon $e_g$ is associated with a vertex if $A(v) \ket\psi = \omega^g \ket\psi$, and an anyon $m_g$ is associated with a plaquette if $B(p) \ket\psi = \omega^g \ket\psi$. The anyonic vacuum corresponds to the stabilizer space of the code, and forms the degenerate ground state space of the model's Hamiltonian, which may be expressed as,
\be \label{ham}
H = - \sum_v A(v) - \sum_p B(p) + \rm{h.c.}.
\ee
This assigns energy to anyons, making them localised quasiparticle excitations.

The creation and movement of the anyons may be achieved by $\sigma^z$ and $\sigma^x$ operations. The operation $\sigma^z$ on spins $1$ or $2$ of a vertex, or $\sigma^z{}^{\dagger}$ on $3$ or $4$, creates an $e_g$ charge at that vertex and an $e_{-g}$ on the other vertex shared by the spin. Similarly, a $\sigma^x$ on spins $2$ or $3$ of a plaquette, or a $\sigma^x{}^{\dagger}$ on $1$ or $4$, creates an $m_g$ flux on that plaquette and an $m_{-g}$ on the other plaquette shared by the spin. Particles can be moved and braided using corresponding strings and loops of the $\sigma^z$ and $\sigma^x$ operations. The commutation relations of these give a phase $\omega^{gh}$ when an $e_g$ anyon is moved clockwise around an $m_h$. The phase $\omega^{-gh}$ is obtained for an anticlockwise braiding. Superpositions of anyon states may be moved using entangling gates \cite{brennen,wootton} or potential wells \cite{pachos}, allowing more complex braiding behaviour.

\section{Encoding of Qudits}\label{encoding}

Logical qudits are stored on the lattice using states of anyon occupancies. The qudits come in two types, denoted $v$ and $p$, stored on pairs of vertices and plaquettes, respectively. We consider a basis for these qudits labelled $\ket{j}_{v/p}$ for $j=0 \ldots d-1$. These are defined in terms of the anyon occupancies of two vertices, $v_1$ and $v_2$, or two plaquettes, $p_1$ and $p_2$, as follows,
\be \label{states}
\ket{j}_v = \ket{e_j}_{v_1}\ket{e_{-j}}_{v_2}, \,\,\,
\ket{j}_p = \ket{m_j}_{p_1}\ket{m_{-j}}_{p_2}.
\ee
The stabilizers are not enforced on the vertices and plaquettes in which the qudits are stored. This opens so-called holes in the code \cite{raussendorf}, bringing the logical states into the stabilizer space. The Hamiltonian (\ref{ham}) must be modified accordingly, with the corresponding plaquette and vertex terms removed. When the anyon quasiparticles are moved, e.g. for braiding, the holes must move with them.

Pauli operators for the qudits are defined in a similar way to the lattice spins,
\be
X = \sum_{j} \ket{j+1}\bra{j}, \,\,\, Z = \sum_{j} \omega^j \ket{j}\bra{j}.
\ee
Certain logical operations follow naturally from the properties of the anyon model. Measurements in the logical $Z$ basis may be performed by measuring the occupancies of the vertices and plaquettes. Logical $Z$ and $X$ rotations may be performed by braiding and fusing with ancillary anyon pairs, respectively. The $X_s$ operation may be performed on the $v$ qudits by creating an $e_1$ and $e_{-1}$, placing the former on $v_1$ and the latter on $v_2$. Similarly, we may perform an $X_p$ operation by placing fluxes on plaquettes. The $Z_s$ operation may be performed on the $v$ qudits by creating an $m_1$ and $m_{-1}$, braiding the $m_1$ around $s_1$ and then annihilating the flux pair. Similarly, we may perform $Z_p$ by braiding a charge pair around $p_1$. Braiding the contents of the $v_1$ of one qudit around those of the $p_1$ of another will implement a controlled-$Z$ gate or its inverse, depending upon the chirality of the operation.

Note from (\ref{states}) that the logical states $\ket{\tilde 0}_{v/p}$ will take the form of Bell pairs. This anyonic entanglement is a natural consequence of the conservation law requiring that anyons are created in pairs. It has been shown that anyonic entanglement is truly non-local, since it may be used to violate Bell's inequalities \cite{iblisdir}. In our case, the entanglement allows the qudits to be stored non-locally, providing fault-tolerance.

\section{Non-topological Operations and Universality}

The logical operations implemented by braiding and fusion do not form a powerful gate set. Further logical operations may be achieved by measuring the spins of the lattice. This allows logical qudits stored on neighbouring plaquettes or vertices to be measured in the $X$ basis. It also allows the preparation of a wide range of logical states. We demonstrate how these may be used as ancillae in protocols implementing various single qudit gates. When $d$ is prime, we prove the universality of the resulting gate set.

\begin{figure}[ht]
\begin{center}
{\includegraphics[width=6cm]{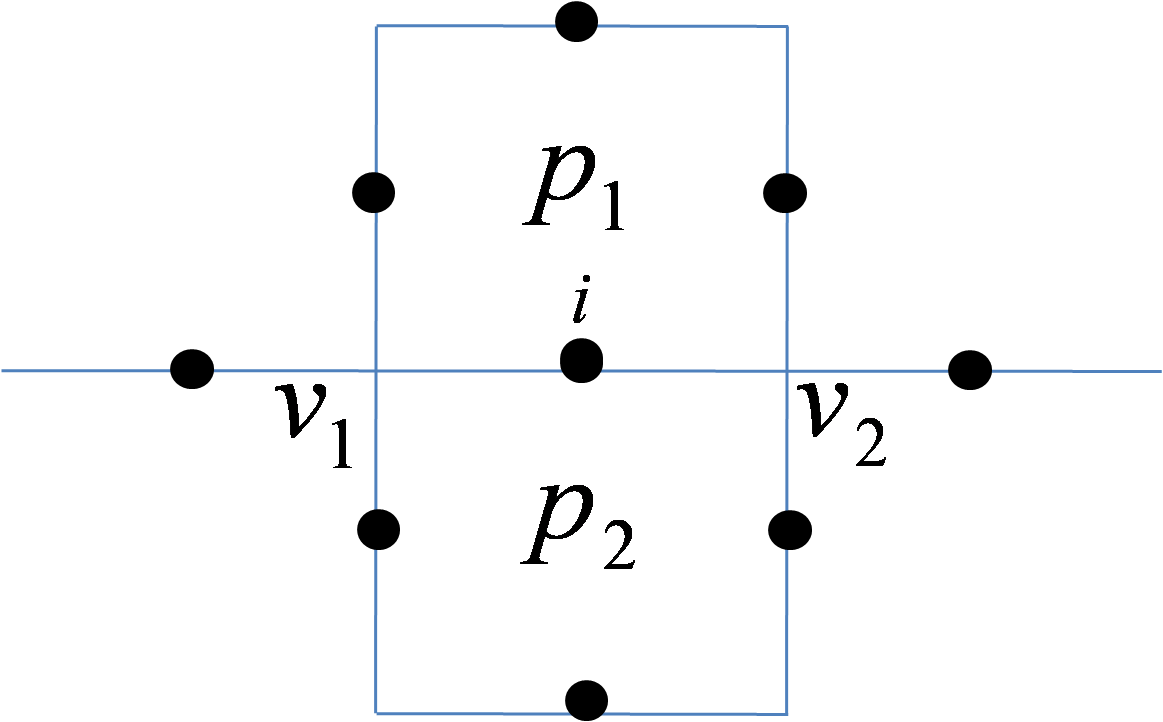} }
\caption{\label{fig2}Measurements of a single spin $i$ effect the surrounding plaquettes and vertices. A $v$-type qudit may be stored in the two vertices and a $p$-type qudit in the two plaquettes. The single spin measurements correspond to either single qudit measurements, or entangling two qudit measurements.}
\end{center}
\end{figure}

Consider two logical qudits, one $v$-type and one $p$-type, stored on the vertices and plaquettes neighbouring a lattice spin, $i$. The vertex $v_1$ is taken to be that for which $i$ is labelled $1$ or $2$ and the plaquette $p_1$ is that for which $i$ is $2$ or $3$ (Fig. \ref{fig2}). The logical $X$ operations then take the following simple form,
\be \nonumber
X_v = \sigma^z_i, \,\,\, X_p = \sigma^x_i.
\ee 
Any projector onto states of $i$ may be decomposed as a sum of the spin Pauli operators, and therefore the logical qudit Pauli operators. Single spin measurements on $i$ therefore correspond to measurements of the logical qubits. In general, these will be in an entangled basis. The exceptions are measurements of $\sigma^z_i$ and $\sigma^x_i$, which provide $X$ basis measurements of the $v$- and $p$-type qudits, respectively.

\begin{figure}[ht]
\begin{center}
{\includegraphics[width=10cm]{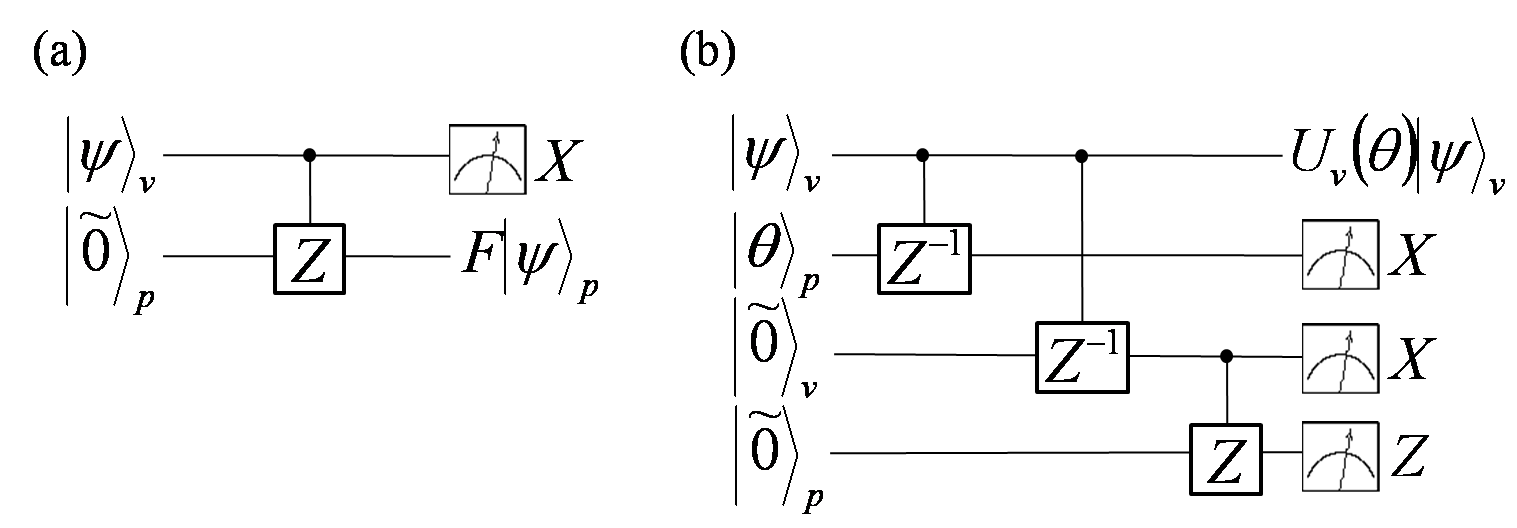} }
\caption{\label{fig3} The circuits to implement (a) a Fourier transform, and (b) the rotation $U(\theta)$. Both of these are implemented on a $v$-type qudit in arbitrary state $\ket{\psi}$. Corresponding circuits exist for $p$-type qudits.}
\end{center}
\end{figure}

The $X$ measurements may be used to implement a Fourier transform, $F$, on a $v$ type qudit in arbitrary state $\ket{\psi}_v=\sum_j c_j \ket{j}_v$. This requires an additional $p$-type qudit in state $\ket{\tilde 0}_p = \sum_k \ket{k}_p / \sqrt{d}$, which can be prepared probabilistically by $X_p$ measurements. Entangling these together with a controlled-$Z$ yields the state,
\be
\Lambda(Z) \ket{\psi}_v \ket{\tilde 0}_p = \sum_{j,k} \omega^{jk} c_j \ket{j}_v \ket{k}_p = \frac{1}{\sqrt{d}} \sum_{l} \ket{{\tilde l}}_v \big( X_p^l F \ket{\psi}_p \big).
\ee
Measuring the $v$ qudit in the $X$ basis then teleports the state to the $p$-type qudit while implementing the Fourier transform $F$. This is up to a Pauli correction, according to the measurement outcome. A corresponding process can be used if the initial state is on a $p$-type qudit. The circuit for the process is given in Fig. \ref{fig3}(a). The Fourier transform allows the application of a controlled-$X$ between two $v$-type qudits, or two $p$-type qudits. This is done by the application of $F^\dagger$ to the target qudit, followed by the controlled-$Z$, and finally followed by $F$.

Single spin measurements may be used to prepare logical ancilla states as follows. First a lattice spin, $i$, is measured in a basis including the state $\ket{\phi} = \frac{1}{\sqrt d}\sum_j e^{i \phi_j} \ket{j}$. The corresponding projector may be expressed,
\be
\ket{\phi}\bra{\phi} = \frac{1}{d}\sum_{j,k} e^{i (\phi_{j+k} - \phi_{j})} \ket{j+k}_i \bra{j} =  \frac{1}{d^2} \sum_{j,k,l} e^{i (\phi_{j+k} - \phi_{j})} \omega^{-jl} X_p^k X_v^l.
\ee
This results in a superposition of the neighbouring $v$ and $p$ qudit states, whose co-efficients depend on those of $\ket{\phi}$. Applying $X_p^\dagger$ to the $p$ qudit and projecting the its state onto $\ket{0}_p$, will leave the $v$ qudit in the state
$$
\frac{1}{\sqrt d} \sum_{j} e^{i (\phi_{j+1} - \phi_{j})} \ket{-\tilde{j}}_v
$$
A Fourier transform using the above method will then transform this into the $p$ qudit
$$
\frac{1}{\sqrt d} \sum_{j} e^{i (\phi_{j+1} - \phi_{j})} \ket{j}_p
$$
Setting $\phi_0 = 0$ and $\phi_j = -\sum_{k=1...j-1} \theta_k$  for $j=1...d-1$, this becomes
\be
\ket{\theta}_p = \sum_{j} e^{i \theta_j} \ket{j}_p,
\ee
which is the required ancilla state.

The ancilla states may be used to perform single qubit rotations of the form,
\be
U_{v,p} (\theta) = \sum_j e^{i \theta_{j}} \ket{j}_{v,p} \bra{j},
\ee 
to $p$-type qudits in arbitrary state $\ket{\psi}_p=\sum_k c_k \ket{k}_p$. To do this, the inverse of a controlled-$X$ is applied between $\ket{\psi}_p$ and an ancilla in state $\ket{\theta}_p$, with the latter as the target. The resulting state is
\be
\sum_{j,k} e^{i \theta_j} c_k \ket{k}_p \ket{j-k}_p = \sum_{j,k} e^{i \theta_{j+k}} c_k \ket{k}_p\ket{j}_p
\ee 
Measuring the second qudit in the $Z$ basis and obtaining the outcome  $j=0$ means the phase gate has been implemented correctly. Otherwise the process may be repeated with the same qudits until the right result is obtained, changing the $\theta_j$'s to correct the erroneous rotations. Each attempt succeeds with a probability of $0.5$, so success will come within a small number of steps. The circuit for the process is given in Fig. \ref{fig3}(a). 

These phase gates, along with the Fourier transform, are able to perform a large number of single qudit rotations. For the case that $d$ is prime it is proven that arbitrary single qudit rotations are possible \cite{clark}. With the entangling gates, this provides universal quantum computation.

\section{Fault-tolerance}

We consider two forms of error in the model: perturbations in the Hamiltonian and thermal error. The former only affect the encoded information if they lift the degeneracy.  Due to the non-local encoding of the qudits, such perturbations must consist of many body operations that either loop around the holes or stretch between them. These become increasingly unlikely as the distance between the holes is increased. Any other perturbations will not commute with the Hamiltonian, and so their effects are efficiently suppressed by the gap.

Thermal errors applied to the spin lattice correspond to the propagation of stray anyons. The creation of these costs energy, and sothe gap natually protects against them. Once created, the movement of stray anyons does not cost further energy. The fault-tolerance of the encoding then comes from the underlying stabilizer code \cite{dennis}, since measuring the syndrome allows stray anyons to be detected and annihilated. If the anyons propagated far enough the annihilation will not correct the error, but consolidate it. For example, consider a $v$-type qudit stored in two holes, $v_1$ and $v_2$, $d$ vertices apart. If the syndrome is measured and a single $e_{-1}$ anyon is found, its antiparticle must reside in one of the holes. It will be assumed that it resides in the hole closest to the $e_{-1}$, since this is the most probable case, and so the $e_{-1}$ will be moved there to annihilate the $e_{1}$ and correct the error. However, if the $e_{1}$ actually resided in the hole furthest from the $e_{-1}$, this action would not correct the error, but actually form a logical $X$ or $X^\dagger$ operation on the qudit. Such a process requires errors to occur on $\> d/2$ spins, and so has probability $O(p^{d/2})$, where $p$ is the probability of a single spin error. Moving the holes far apart therefore exponentially suppresses $X$ errors. A corresponding argument holds for $p$-type qudits.

The occurrence of $Z$ errors requires the braiding of flux anyons around holes. Since the boundary of a hole consists of four spins, only two spin errors are required to fool the stabilizer and cause logical errors. This is because such errors form half of a loop around a hole, and attempts at error correction may complete this loop by mistake. The solution to this is to use a repetition code, which emerges naturally from the anyon models. To do this for $v$-type qudits, two additional holes are introduced. The first is on a vertex neighboring $v_1$, denoted $v_1'$. The shared spin between these is projected onto the state $\ket{0}$, corresponding to the identity element of $Z_d$. To do this, the spin may be measured in the $\sigma^z$ basis. If the outcome is $\ket{g}$, the operation $A^{\pm g} (v_1')$ may be applied, where the sign depends upon the orientation of the link. This then maps the spin state to $\ket{0}$ without creating flux anyons. The second additional hole is on $v_2'$, neighbouring $v_2$, and the shared spin is similarly projected to $\ket{0}$. The projections cause the logical qudit to be stored in all four holes, and so it becomes further delocalised. Repeated applications of the procedure allow the qudit to be stored in $2N$ holes, arranged in two rows. The application of a logical $Z$ requires flux anyons to braid around entire rows, rather than single holes. For this to occur in error, an anyon pair must propagate around at least half of the loop, which has probability $O(p^{d/2})$. The addition of holes therefore exponentially suppresses $Z$ errors. The probablility of $X$ errors, however, increases since there are more holes that stray anyons may fall into. This increase is polynomial, and so may easily be compensated for by increasing the distance between the rows.

This scheme is called a repetition code since the $X$ basis measurements can be performed using any two vertices from different rows, and with loops around single holes affected only the $X$-basis measurements using that hole. Measurement of the syndrome is then equivalent to majority voting with these. A corresponding scheme can be used to store $p$-type qudits on more holes, though the projections in that case are to the state $\ket{\tilde 0}$ using $B^{\pm g} (p_1')$ operations.

The preparation of the logical ancilla states requires the holes to be neighbouring, and so prone to errors. The preparation should therefore occur on quickly, on timescales for which the probability of error is small. In addition to this, errors in preparation can be suppressed by using purification schemes \cite{bravyi}.

\section{Application to Other Models}

The methods developed above have a straightforward generalisation to all other quantum double models. These are similarly realized on spin lattices, with stabilizers defined on each plaquette and vertex \cite{kitaev}. Flux quasiparticles are associated with non-trivial eigenvalues of the plaquette stabilizers and charge quasiparticles are associated with those of the vertices. For the abelian quantum double models, those based upon an abelian group, these quasiparticles are gauge invariant and may be directly identified with anyons. For the non-abelian models the quasiparticles are gauge dependent the identification is not so straightforward. Even so, it is known how the gauge invariant anyons may be formed from these \cite{brennen}.

Logical information may be stored by not enforcing stabilizers, and so creating holes. For the abelian models, the logical information corresponds the anyon states stored within the holes. For the non-abelian models, the encoding could be expressed either in terms of the anyons, or the underlying flux and charge quasiparticles \cite{mochon}. The separation between the holes provides fault-tolerance, with errors suppressed by distance. The usage of additional holes will also provide protection for the abelian models and in some cases for the non-abelian, but the full generalization is not known.

For any quantum double models, projections onto lattice spins may be decomposed into creation operators for quasiparticle pairs. Spin measurements may then be used to measure logical states stored on neighboring plaquettes and vertices, and to create ancilla states. These will then increase the computational power of the model from that possible by braiding and fusion alone. Note that, for non-abelian models, it may not be possible to measure $v$-type and $p$-type qudits separately, corresponding to the $X$ measurements above. It is likely that only entangling measurements between a $v$- and a $p$-type qudit will be possible in general.

The application of the methods beyond quantum double models is speculative. We must therefore use general principles to guide us. In any physical realization of anyons, there are two types of entanglement to consider: that of the underlying physical medium and that of the anyons. Any controlled reduction in the former seems to yield increases in the latter. For example, the state with vacuum everywhere has no anyonic entanglement, but requires a maximally entangled state of the physical medium. Conversely, a separable state of the medium results in indefinite anyon occupancies across the system. The corresponding superpositions lead to highly entangled states of anyons. Localised measurements on the medium will lower its entanglement. Anyonic entanglement should then arise as a consequence, which could be used to prepare logical ancilla states and increase computational power.

The method can also be applied to non-anyonic stabilizer codes. Consider the following code, defined on six spin-$1/2$ particles.
\bq
S_1 &=& \sigma^z_1 \sigma^z_2, \,\, S_2 = \sigma^z_2 \sigma^z_3, S_3 = \sigma^z_4 \sigma^z_5, \,\, S_4 = \sigma^z_5 \sigma^z_6, \\ \nonumber
S_5 &=& \sigma^x_1 \sigma^x_2 \sigma^x_3 \sigma^x_4 \sigma^x_5 \sigma^x_6, S_6 = \sigma^z_1 \sigma^z_3 \sigma^z_4 \sigma^z_6.
\eq
A qubit may be stored in holes by not enforcing stabilizers $S_1$ and $S_2$. The logical state $\ket{0}$ is identified with a $+1$ eigenvalue for both $S_1$ and $S_2$. The state $\ket{1}$ is associated with a $-1$ eigenvalue. The logical Pauli operators are $Z = S_1 \, {\rm or} \, S_2$ and $X = \sigma^x_2$. Alternatively, a qubit may be stored by not enforcing $S_2$ and $S_3$. The Pauli operators in this case are $Z = S_2 \, {\rm or} \, S_3$ and $X = \sigma^x_3 \sigma^x_4$. In the former case $X$ is provided by a single spin operation, and so the information stored in this way is easily accessible. This allows the possibility of creating states by single spin measurements, as with anyons. In the latter case the $X$, like the $Z$'s, is a two spin operation. This protects the stored information, since single spin operations are not sufficient to cause a logical error.

With anyonic codes, the separation of holes means that logical qudits can be given arbitrary protection. In the above example the protection is limited to single spin errors. This limitation is also found when information is stored in holes of other stabilizer codes, such as the Steane and Shor codes. Of course this is not suprising, since these codes were designed to have information stored in the stabilizer space, rather than in holes. It would be interesting to know whether topological codes are the only ones for which holes provide arbitrary protection, or whether others are possible.

\section{Experimental Realizations}

The use of simple abelian models means that our work is in tune with current experimental proposals, as well as actual experiments. One promising possibility is Josephson-junctions, for which the means to produce the Hamiltonians for $D(Z_d)$ models has been well studied \cite{doucot,gershenson}, along with other models such as the non-abelian $D(S_3)$. Polar molecules have also been considered as a means to produce the Hamiltonians of quantum double models, with a complete toolbox of interactions proposed \cite{zoller}. Topological states may also be produced without a Hamiltonian. Experiments with these have been performed, demonstrating anyonic braiding \cite{pachos07,lu}. These systems, though small, could also be used to demonstrate the preparation of ancilla state described here.

\section{Conclusion}

We have studied a class of anyon models, realizable on spin lattice. These are abelian, and so not expected to possess any serious computational power. We show that the addition of single spin measurements increases the computational power significantly. This is because they may be expressed in terms of the logical states as measurements in entangled bases. This additional resource provably provides a universal gate set for many of these models, and likely for most others as well. We also comment on the fault-tolerance, naturally provided by the Hamiltonian and stabilizers of the topological models on which the schemes are based. This work also gives an insight into the relationship between abelian and non-abelian models. The repetition code used, for example, bears striking similarities to the encoding found in the non-abelian charges of the $D(S_3)$. Elaborations on these points may be found in further works by the authors \cite{wootton2,wootton3}.

\end{document}